\documentclass[a4paper,11pt]{article}
\usepackage{jcappub} 
\usepackage{amsmath}
\usepackage{lineno}

\title{The formalism of energy conservation during particle decay in the Kerr spacetime}

\author{Shu-Rui Zhang$^{1,2}$\,\href{https://orcid.org/0000-0003-0368-384X}{\includegraphics[width=0.8em]{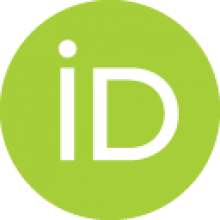}}}
\author{Remo Ruffini$^{1,2,3,4}$\,\href{https://orcid.org/0000-0003-0829-8318}{\includegraphics[width=0.8em]{orcid-ID.png}}}%
\affiliation{$^{1}$ICRANet, Piazza della Repubblica 10, I-65122 Pescara, Italy}
\affiliation{$^{2}$University “G. d’Annunzio” of Chieti-Pescara, I-66100 Chieti, Italy}
\affiliation{$^{3}$ICRA, Dipartimento di Fisica, Sapienza Universit$\grave{a}$ di Roma, Piazzale Aldo Moro 5, I-00185 Roma, Italy}
\affiliation{ $^{4}$INAF, Viale del Parco Mellini 84, 00136 Rome, Italy}

\emailAdd{zhangsr@mail.ustc.edu.cn (SRZ)}
\emailAdd{ruffini@icra.it (RR)}

\abstract{
We derive a compact, covariant expression for the relative Lorentz factor of two particles in curved spacetime and apply it to particle decay in Kerr spacetime. This allows us to show that energy conservation in the local center-of-mass frame requires the rest-mass loss of the parent particle to be converted into kinetic
energy of the decay products. We verify this relation analytically and confirm it with high-precision reconstructions of three representative Penrose-process examples, for which both equivalent conservation formulas are satisfied to machine precision. These results clarify the local kinematics underlying energy extraction from rotating black holes and show that mass loss is not optional but is required for the decay products to separate.}

\begin{document}
\maketitle
\flushbottom

\section{Introduction}
\label{sec:intro}
The classic textbook on general relativity and astrophysics, MTW \cite{Misner:1973prb}, introduces the concept of energy extraction from black holes, based on Penrose's earliest work \cite{1969NCimR...1..252P}. It describes an advanced civilization sending a vehicle loaded with garbage into the ergosphere region of a Kerr spacetime. The vehicle dumps the garbage into the black hole and then returns empty, potentially extracting rotational energy from the black hole. Although this description is heuristic, it is not entirely accurate, as it implicitly assumes mass conservation. While, we will show that during the Penrose process, mass loss is necessary to generate kinetic energy in the center-of-mass frame.

For more than fifty years, the Penrose process \cite{1969NCimR...1..252P,1971NPhS..229..177P} has raised a subtle question concerning the relation between the rest mass of the progenitor particle before splitting and the much smaller rest masses of the two fragments produced in the process. This issue can already be traced back to the first explicit computation of the Penrose process, presented by Ruffini and Wheeler in Fig.~2 of the paper by Christodoulou \cite{1970PhRvL..25.1596C}. Wald further derived strict limits on the energies obtainable from the Penrose process, demonstrating that one cannot obtain energies significantly greater than those achievable through a similar breakup process in flat spacetime \citep{1974ApJ...191..231W}. In our recent work \cite{2025PhRvL.134h1403R}, we showed that three representative examples with different fragment masses and Lorentz factors are possible. Here, by reexamining these examples in the center-of-mass frame, we uncover the underlying identity that resolves this apparent ambiguity. An infinite family of Penrose processes is in fact possible, with different values of the fragment masses and of the progenitor mass $m_0$. However, in each individual process, the rest-mass energy of the progenitor is exactly equal to the total energy of the debris when both their rest masses and their kinetic energies are evaluated in the center-of-mass frame. Thus, the correct conservation law is not a conservation of rest mass alone, but a conservation of total mass-energy in the local rest frame of the system.

\section{The formula for the relative velocity of two particles}

Here we will show a concise formula for the relative velocity of two particles in terms of the four-velocity of the two particles in curve spacetime. It is useful in many physical scenarios, including verify the energy conservation of the Penrose process.

Consider the collision or fission of two particles. Namely, let the masses of the two particles be $m_1$ and $m_2$, located at the same spacetime point but with different four-velocities $u_1$ and $u_2$. The four-momentum of each particle can be expressed as:
\begin{gather}
p_1^\alpha = m_1 u_1^\alpha, \quad p_2^\alpha = m_2 u_2^\alpha.
\label{eq:p12}
\end{gather}
Thus, the total four-momentum of the two particles is:
\begin{gather}
(p_{\text{tot}})^\alpha = p_1^\alpha + p_2^\alpha = m_1 u_1^\alpha + m_2 u_2^\alpha.
\label{eq:pt}
\end{gather}
The center-of-mass energy of the two particles $E_{\text{cm}}$ is then given by (e.g.,\cite{2011PhRvD..83h4041H}):
\begin{gather}
E_{\text{cm}}^2 = -(p_{\text{tot}})^\alpha (p_{\rm tot})_\alpha=m_1^2+m_2^2-2m_1 m_2\ g_{\rm \alpha \beta} u_1^\alpha u_2^\beta.
\label{eq:Ecm1}
\end{gather}

On the other hand, consider the local reference frame in which particle 1 is at rest. In this frame, the four-velocities of particle 1 and particle 2 are given by:
\begin{gather}
u_1 = (1, 0, 0, 0), \quad u_2 = (\gamma_{\rm r}, \gamma_{\rm r} \mathbf{v}_{\rm r})
\label{eq:u12}
\end{gather}
where \(\gamma_{\rm r} = \frac{1}{\sqrt{1 - v_{\rm r}^2}}\) is the relative gamma factor between the two particles. According to the equivalence principle, the center-of-mass energy of the two
particles follows special relativity, namely, it satisfies (e.g., \citep{landau1975classical}):
\begin{gather}
E_{\text{cm}}^2 = m_1^2 + m_2^2 + 2m_1m_2 \gamma_{\rm r}.
\label{eq:Ecm2}
\end{gather}
When \(\gamma_{\rm r} = 1\), i.e., when there is no relative velocity between the two particles, the expression is simplified to the obvious form:
\begin{gather}
E_{\text{cm}} = m_1 + m_2
\label{eq:m1+m2}
\end{gather}

It is worth noting that the center-of-mass energy of the two particles (collision or fission) is determined solely by their masses and relative velocity, regardless of the spacetime property and the reference frame used. Therefore, Equation (\ref{eq:Ecm2}) is always satisfy and equivalent to Equation (\ref{eq:Ecm1}). By comparing the right-hand sides of these two equations, an important identity is obtained.:
\begin{gather}
- g_{\rm \alpha \beta} u_1^\alpha u_2^\beta = \gamma_{\rm r} = \frac{1}{\sqrt{1 - v_{\rm r}^2}}.
\label{eq:identity}
\end{gather} 
If the two particles have no relative velocity, \(u_1 = u_2 = u\), the above equation is simplified to the well-known four-velocity normalization equation:
\begin{gather}
- g_{\rm \alpha \beta} u^\alpha u^\beta = 1.
\label{eq:normalization}
\end{gather}

\section{Energy conservation during particle fission in Kerr spacetime}

In Kerr spacetime, when particle 0 (with mass \( m_0 \)) decays into particle 1 (with mass \( m_1 \)) and particle 2 (with mass \( m_2 \)), momentum conservation in each direction must be satisfied, i.e.,
\begin{subequations}\label{eq:4eqs}
    \begin{align}   
    p_{\rm t0}&= p_{\rm t1} + p_{\rm t2},\label{subeq:4eqsa}\\
    p_{\rm r0}&= p_{\rm r1} + p_{\rm r2},\label{subeq:4eqsb}\\
    p_{\rm \theta 0}&= p_{\rm \theta 1} + p_{\rm \theta 2},\label{subeq:4eqsc}\\
    p_{\rm \phi 0}&= p_{\rm \phi 1} + p_{\rm \phi 2},\label{subeq:4eqsd}
    \end{align}
\end{subequations}
where \( p_{\alpha i} \) represents the $\alpha$ component of four-momentum of particle $i$ ($i$=0, 1, and 2). In this process, we want to verify that the sum of the two masses \( m_1 + m_2 \) is necessarily less than \( m_0 \); and the mass loss is converted into kinetic energy in the center-of-mass frame. Namely, we want to verify:
\begin{gather}
m_0 = m_1 \gamma_{01} + m_2 \gamma_{02},
\label{eq:massloss}
\end{gather}
where $\gamma_{01}$ is relative gamma factor between particle 0 and 1, and $\gamma_{02}$ is relative gamma factor between particle 0 and 2. If $m_0 > m_1 + m_2$, then we naturally have $\gamma_{01}, \gamma_{02} > 1$, indicating that the two particles have a relative velocity and separate.

Using the previously derived Equation (\ref{eq:identity}), the right-hand side of Equation (\ref{eq:massloss}) can be written as:
\begin{gather}
m_1 \gamma_{01} + m_2 \gamma_{02} = -g_{\rm \alpha \beta} \left(m_1 u_0^{\alpha} u_1^{\beta} + m_2 u_0^{\alpha} u_2^{\beta} \right).
\label{eq:masslossright}
\end{gather}
Note that \( u_{\rm i}^{\alpha} \) represents the four-velocity of the particle i, which is expressed in terms of the conserved quantities of the particle's motion. Similarly, the momenta of the particles in each direction in Equation (\ref{eq:4eqs}) can also be expressed in terms of the conserved quantities of the particle's motion. Thus, substituting Equation (\ref{eq:4eqs}) into the right-hand side of Equation (\ref{eq:masslossright}), after a lengthy but straightforward calculation (details in Appendix), the right-hand side of Equation (\ref{eq:masslossright}) is found to equal $m_0$, i.e.,
\begin{gather}
-g_{\rm \alpha \beta} \left(m_1 u_0^{\alpha} u_1^{\beta} + m_2 u_0^{\alpha} u_2^{\beta} \right) = m_0,
\label{eq:m0}
\end{gather}
where $g_{\rm \alpha \beta}$ is the Kerr metric.
Therefore, we have proved that Equation (\ref{eq:massloss}) holds generally in Kerr spacetime. So we also have the following relations:
\begin{gather}
m_0 = m_1 \gamma_{01} + m_2 \gamma_{02} = \sqrt{m_1^2 + m_2^2 + 2 m_1 m_2 \gamma_{12}},
\label{eq:Ecm}
\end{gather}

Moreover, it is straightforward to extend the above results to a more general case, where particle 0 decays into $N$ particles ($N\geqslant 2$). The corresponding momentum conservation equations for each direction thus are: 
\begin{gather}
p_{\rm \alpha 0} = \sum_{i=1}^{N} p_{\alpha i}\ (,{\rm where}\ \alpha = t, r, \theta, \phi).
\label{eq:genmonentum}
\end{gather}
The corresponding formula (\ref{eq:massloss}) generalizes to:
\begin{gather}
m_0 = \sum_{i=1}^{N} m_i \gamma_{0i},
\label{eq:genmassloss}
\end{gather}
where $\gamma_{0i}$ is relative gamma factor between particle 0 and $i$, and it satisfies the Equation (\ref{eq:identity}) in any curve spacetime. Additionally, at least 
\begin{gather}
\begin{cases}
N-1, & N > 2 \\
2, & N = 2
\end{cases}
\end{gather}
of these relative gamma factors are strictly lager than 1. In other words, at most one of the decay particles can follow the trajectory of particle 0, thus the corresponding relative gamma factor equals 1. Of course, any two of the decay particles must have a relative velocity (i.e., $\gamma_{ij}>1, {\rm where}\ i,j \geqslant 1 \ \&\ i\neq j$ ) to ensure that the $N$ particles are separated from each other.

\section{Examples}
\label{sec:examples}

Now, we present illustrative examples to show the formalism of energy conservation
during particle decay in the Kerr spacetime. We use the three examples in Table 1
in \cite{2025PhRvL.134h1403R}. Since the original data are not sufficiently
accurate, they cannot be directly used to validate the energy conservation
formula. For example, in the case of $r=1.2M$, the actual four-velocity is
$u_2=(769.2123074205166,0,0,349.6387076994172)$, whose large magnitude leads
to significant inaccuracies if numerical precision is not properly handled.
Therefore, we have recovered all quantities with higher numerical precision and
verified that all cases in the \cite{2025PhRvL.134h1403R} paper satisfy the
energy conservation relation, as expected. Here we summarize them below.

For all three cases, we take
\begin{equation}
    \hat{E}_0=1,\qquad \hat{\phi}_1=-19.434,\qquad
    \frac{\mu_2}{\mu_1}=0.78345 .
\end{equation}
The recovered masses and conserved quantities are listed in
Table~\ref{tab:examples-conserved-quantities}. Here the masses of particles 1
and 2 are normalized by the mass of particle 0.

\begin{table}[htbp]
\centering
\caption{Recovered masses and conserved quantities for the three examples.}
\label{tab:examples-conserved-quantities}
\begin{tabular}{ccccc}
\hline
$r/M$ & Particle & $m_i/m_0$ & $\hat{E}_i$ & $\hat{\phi}_i$ \\
\hline
$1.2$ & 0 & $1$ & $1$ & $2.1127016653792583$ \\
      & 1 & $0.02090057578214192$ & $-6.9355825766935695$ & $-19.434$ \\
      & 2 & $0.016374556096519087$ & $69.92297455201749$ & $153.82911391807868$ \\
\hline
$1.5$ & 0 & $1$ & $1$ & $2.267949192431123$ \\
      & 1 & $0.021818440539206844$ & $-3.520629285604805$ & $-19.434$ \\
      & 2 & $0.017093657240441602$ & $62.99498261735006$ & $157.4834880566801$ \\
\hline
$1.9$ & 0 & $1$ & $1$ & $2.4557701793438658$ \\
      & 1 & $0.0246951592273348$ & $-0.5007583467069353$ & $-19.434$ \\
      & 2 & $0.01934742249665545$ & $52.32564220279732$ & $151.73576243943492$ \\
\hline
\end{tabular}
\end{table}

The corresponding four-velocities are given in
Table~\ref{tab:examples-four-velocities}. With these recovered high-precision
values, the four-velocities of particles 0, 1, and 2 are normalized to
$u^\alpha u_\alpha=-1$ to machine precision.

\begin{table}[htbp]
\centering
\caption{Recovered four-velocities $u^\alpha=(u^t,u^r,u^\theta,u^\phi)$.}
\label{tab:examples-four-velocities}
\begin{tabular}{ccc}
\hline
$r/M$ & Particle & $u^\alpha$ \\
\hline
$1.2$ & 0 & $(14.63743060919755,\,0,\,0,\,6.45497224367902)$ \\
      & 1 & $(97.69685545946016,\,0,\,0,\,34.91739263776793)$ \\
      & 2 & $(769.2123074205166,\,0,\,0,\,349.6387076994172)$ \\
\hline
$1.5$ & 0 & $(6.237604307034004,\,0,\,0,\,2.3094010767585007)$ \\
      & 1 & $(39.10312976391187,\,0,\,0,\,7.135310476774361)$ \\
      & 2 & $(314.99607834912354,\,0,\,0,\,125.99525655029342)$ \\
\hline
$1.9$ & 0 & $(3.799518935225643,\,0,\,0,\,1.139975946761282)$ \\
      & 1 & $(21.754607503430847,\,0,\,0,\,0.6120099457999526)$ \\
      & 2 & $(168.6160231195753,\,0,\,0,\,58.14016124863621)$ \\
\hline
\end{tabular}
\end{table}

Using Eq.~(2.7), we compute the relative Lorentz factors
$\gamma_{01}$, $\gamma_{02}$, and $\gamma_{12}$. The numerical results are
summarized in Table~\ref{tab:examples-gamma-check}. They explicitly verify both
forms of the energy conservation relation,
\begin{equation}
    m_0=m_1\gamma_{01}+m_2\gamma_{02},
    \label{eq1}
\end{equation}
and
\begin{equation}
    m_0=\sqrt{m_1^2+m_2^2+2m_1m_2\gamma_{12}} .
    \label{eq2}
\end{equation}

\begin{table}[htbp]
\centering
\caption{Relative Lorentz factors and numerical checks of energy conservation.}
\label{tab:examples-gamma-check}

\begin{tabular}{cccc}
\hline
$r/M$ & $\gamma_{01}$ & $\gamma_{02}$ & $\gamma_{12}$ \\
\hline
$1.2$
& $23.926821882946435$
& $30.530027382906207$
& $1459.9431683064818$ \\
$1.5$
& $22.92060813036612$
& $29.24523800573388$
& $1339.6073975118115$ \\
$1.9$
& $20.25165172907347$
& $25.837148902930664$
& $1045.4600127403396$ \\
\hline
\end{tabular}

\vspace{0.5em}

\begin{tabular}{ccc}
\hline
$r/M$
& $m_1\gamma_{01}+m_2\gamma_{02}$
& $\sqrt{m_1^2+m_2^2+2m_1m_2\gamma_{12}}$ \\
\hline
$1.2$
& $0.9999999999999951\,m_0$
& $1.0000000000000001\,m_0$ \\
$1.5$
& $1.000000000000005\,m_0$
& $0.9999999999999994\,m_0$ \\
$1.9$
& $0.9999999999999969\,m_0$
& $0.9999999999999999\,m_0$ \\
\hline
\end{tabular}

\end{table}

Therefore, for all three radii, the high-precision reconstructed solutions
provide self-consistent local splitting kinematics. The masses, conserved
quantities, and four-velocities of particles 0, 1, and 2 are all explicitly
listed, and both energy conservation formulas are numerically satisfied to
machine precision.

\section{Discussion and conclusions}
\label{sec:concl}
We verified during particle decay in Kerr spacetime (e.g., Penrose process), the sum of the two masses, $m_1 + m_2$, cannot be equal to the initial particle mass $m_0$ in this process. If the masses remain equal, the solution of trajectories of $m_1$ and $m_2$ always coincide, and they do not separate, meaning that these two particles never have a relative velocity. 

Actually, the covariant identity Equation (\ref{eq:genmonentum}) and Equation (\ref{eq:genmassloss}) shows that any deficit between the progenitor rest mass and the sum of the fragment rest masses is exactly accounted for by the kinetic energy of the debris. Hence, when the fragments separate, rest-mass loss is not an additional assumption but a direct consequence of four-momentum conservation. The high-precision examples further demonstrate that this conclusion is not a numerical artifact of a special case. For the three representative Kerr trajectories considered here, the reconstructed masses, conserved quantities, and four-velocities satisfy both Equation (\ref{eq1}) and Equation (\ref{eq2}) to machine precision. From a physical perspective, mass loss is necessary to generate kinetic energy in the center of the mass frame, allowing the two particles to have a relative velocity and separate. The result therefore removes an apparent
ambiguity in the particle-splitting interpretation of the Penrose process and gives a compact local criterion for physically admissible black-hole energy extraction events.

\appendix
\section{Appendix}
\label{sec:AppendixA}
In this Appendix, we provide a detailed derivation of Equation~(\ref{eq:massloss}), 
and the corresponding Mathematica code is also available (see \href{https://github.com/zhangsr666/Energy-Conservation-in-Curved-Spacetime}{link}).

In Boyer--Lindquist coordinates $(t,r,\theta,\phi)$,
a Kerr black hole with spin parameter $a$ and mass $M$
is described by the metric $g_{\alpha\beta}$,
which can be written in matrix form as
\begin{equation}
g_{\alpha\beta}
=
\begin{pmatrix}
-\left(1-\dfrac{2Mr}{\Sigma}\right)
& 0
& 0
& -\dfrac{2Mar\sin^2\theta}{\Sigma}
\\[0.8em]
0
& \dfrac{\Sigma}{\Delta}
& 0
& 0
\\[0.8em]
0
& 0
& \Sigma
& 0
\\[0.8em]
-\dfrac{2Mar\sin^2\theta}{\Sigma}
& 0
& 0
& \left(r^2+a^2+\dfrac{2Ma^2r\sin^2\theta}{\Sigma}\right)\sin^2\theta
\end{pmatrix},
\end{equation}
where
\begin{equation}
\Sigma \equiv r^2+a^2\cos^2\theta,
\qquad
\Delta \equiv r^2-2Mr+a^2 .
\end{equation}

For a particle $i$ $(i=0,1,2)$ with rest mass $m_i$
moving in the Kerr spacetime,
the conserved quantities are the specific energy $\mathrm{E}_i$,
the axial component of the specific angular momentum $\mathrm{Lz}_i$,
and the Carter constant $\mathrm{Q}_i$.
The four-velocity $u_i^\alpha=\mathrm{d}x^\alpha/\mathrm{d}\tau$
is given by
\begin{subequations}\label{eq:4u}
    \begin{align}   
    \Sigma\,u_i^t&=\frac{(r^2+a^2)\left[\mathrm{E}_i(r^2+a^2)-a\,\mathrm{Lz}_i\right]}{\Delta}-a\left(a\,\mathrm{E}_i\sin^2\theta-\mathrm{Lz}_i\right),\label{subeq:4ua}\\
    \Sigma\,u_i^r&=\pm\sqrt{\mathcal{R}_i(r)},\label{subeq:4ub}\\
    \Sigma\,u_i^\theta&=\pm\sqrt{\Theta_i(\theta)},\label{subeq:4uc}\\
    \Sigma\,u_i^\phi&=\frac{a\left[\mathrm{E}_i(r^2+a^2)-a\,\mathrm{Lz}_i\right]}{\Delta}-\left(a\,\mathrm{E}_i-\frac{\mathrm{Lz}_i}{\sin^2\theta}\right),\label{subeq:4ud}
    \end{align}
\end{subequations}
where 
\begin{equation}
\mathcal{R}_i(r)
\equiv
\left[
a\,\mathrm{Lz}_i
-\mathrm{E}_i\left(a^2+r^2\right)
\right]^2
-
\left[
(\mathrm{Lz}_i-a\,\mathrm{E}_i)^2
+r^2
+\mathrm{Q}_i
\right]
\left[
a^2+r(r-2M)
\right] .
\end{equation}
and 
\begin{equation}
\Theta_i(\theta)
\equiv
a^2(\mathrm{E}_i^2-1)\cos^2\theta
-\mathrm{Lz}_i^{\,2}\cot^2\theta
+\mathrm{Q}_i .
\end{equation}
Note that, as a result of the collision or separation of the two particles, the radial components $u_1^r$ and $u_2^r$ have opposite signs, and the polar components $u_1^\theta$ and $u_2^\theta$ also have opposite signs.

Since
\begin{equation}
p_{\alpha i} \equiv m_i\, g_{\alpha\beta}\, u_i^{\beta} ,
\end{equation}
the momenta of the particles in each direction can also be expressed in terms of the conserved quantities of the particle's motion:
\begin{subequations}\label{eq:4mon}
    \begin{align}   
    p_{t i}&=- m_i \mathrm{E}_i,\label{subeq:4mona}\\
    p_{r i}&=\pm m_i \frac{\sqrt{\mathcal{R}_i(r)}}{\Delta},\label{subeq:4monc}\\
    p_{\theta i}&=\pm m_i \sqrt{\Theta_i(\theta)},\label{subeq:4mond}\\
    p_{\phi i}&= m_i \mathrm{Lz}_i,\label{subeq:4monb}
    \end{align}
\end{subequations}
The radial components $p_{r 1}$ and $p_{r 2}$ as well as the polar components $ p_{\theta 1}$ and $ p_{\theta 2}$ also have opposite signs.

We then expand the right-hand side of Equation~(\ref{eq:masslossright}), i.e.,
\begin{gather}
\begin{aligned}
&-g_{\alpha\beta}\!\left(m_1 u_0^{\alpha} u_1^{\beta}
+ m_2 u_0^{\alpha} u_2^{\beta}\right) \\[0.5em]
&= -\frac{1}{
\Delta^2
\Sigma^3
}
\Bigg\{
-\frac{1}{4}
\Delta
\sqrt{\mathcal{R}_0}
\Big(
m_1 \sqrt{\mathcal{R}_1}
-m_2 \sqrt{\mathcal{R}_2}
\Big)
\left(\cos (2 \theta ) a^2+a^2+2 r^2\right)^2
\\[0.6em]
&\quad
+\frac{1}{4}
\Delta^2
\sqrt{\Theta_0}
\Big(
m_1 \sqrt{\Theta_1}
+m_2 \sqrt{\Theta_2}
\Big)
\left(\cos (2 \theta ) a^2+a^2+2 r^2\right)^2
\\[0.6em]
&\quad
+2 a M r
\Big[
a^2 (\mathrm{Lz}_1m_1+\mathrm{Lz}_2m_2)\cos^2\theta
+2 a M (\mathrm{E}_1m_1+\mathrm{E}_2m_2) r \sin^2\theta
-(\mathrm{Lz}_1m_1+\mathrm{Lz}_2m_2)(2M-r)r
\Big]
\\
&\qquad\times
\Big[
-2 \mathrm{E}_0 r^2 a^2
+\mathrm{E}_0\!\Delta a^2 \sin^2\theta
+2 \mathrm{Lz}_0 M r a
-\mathrm{E}_0(r^4+a^4)
\Big]
\\[0.6em]
&\quad
+2 a M r
\left(
a^2 \mathrm{Lz}_0\cot^2\theta
+\mathrm{Lz}_0 r(r-2M)\csc^2\theta
+2 a \mathrm{E}_0 M r
\right)\sin^2\theta
\\
&\qquad\times
\Big[
-2 (\mathrm{E}_1m_1+\mathrm{E}_2m_2) r^2 a^2
+(\mathrm{E}_1m_1+\mathrm{E}_2m_2)
\Delta a^2 \sin^2\theta
\\
&\qquad\qquad
+2 M (\mathrm{Lz}_1m_1+\mathrm{Lz}_2m_2) r a
-(\mathrm{E}_1m_1+\mathrm{E}_2m_2)(r^4+a^4)
\Big]
\\[0.6em]
&\quad
+\left((2M-r)r-a^2\cos^2\theta\right)
\Big[
-\mathrm{E}_0(a^2+r^2)^2
+a^2\mathrm{E}_0\!\Delta \sin^2\theta
+2 a \mathrm{Lz}_0 M r
\Big]
\\
&\qquad\times
\Big[
-2 (\mathrm{E}_1m_1+\mathrm{E}_2m_2) r^2 a^2
+(\mathrm{E}_1m_1+\mathrm{E}_2m_2)
\Delta a^2 \sin^2\theta
\\
&\qquad\qquad
+2 M (\mathrm{Lz}_1m_1+\mathrm{Lz}_2m_2) r a
-(\mathrm{E}_1m_1+\mathrm{E}_2m_2)(r^4+a^4)
\Big]
\\[0.6em]
&\quad
+\left(
a^2 \mathrm{Lz}_0\cot^2\theta
+\mathrm{Lz}_0 r(r-2M)\csc^2\theta
+2 a \mathrm{E}_0 M r
\right)
\\
&\qquad\times
\left(
a^2 (\mathrm{Lz}_1m_1+\mathrm{Lz}_2m_2)\cot^2\theta
-(\mathrm{Lz}_1m_1+\mathrm{Lz}_2m_2)(2M-r)r\csc^2\theta
+2 a M (\mathrm{E}_1m_1+\mathrm{E}_2m_2) r
\right)
\\
&\qquad\times
\left(
2 a^2 M r \sin^4\theta
+(a^2+r^2)\Sigma\sin^2\theta
\right)
\Bigg\}.
\label{eq:verify}
\end{aligned}
\end{gather}
From Equations~(\ref{eq:4eqs}) and (\ref{eq:4mon}), 
it follows that the above expression simplifies to $m_0$, i.e., Equation~(\ref{eq:m0}).

\acknowledgments

We thank Professor R. Kerr, Professor R. Penrose, Professor Y. Wang, and Professor J.A. Rueda for discussions on this work.



\bibliographystyle{JHEP}
\bibliography{biblio.bib}

\end{document}